\documentstyle[psfig]{iopart}
\begin{document}
\newcommand{\be}{\begin{equation}}
\newcommand{\ee}{\end{equation}}
\newcommand{\bq}{\begin{eqnarray}}
\newcommand{\eq}{\end{eqnarray}}
\def\qsqrt{{\sqrt{2} \kern-1.2em ^4}}
\def\id{{\rm 1\kern-.34em 1}}
\def\RR{{\rm I\kern-.20em R}}
\def\CC{{\rm I\kern-.20em C}}
\def\NN{{\rm I\kern-.20em N}}
\def\up{\uparrow}
\def\dwn{\downarrow}
\def\PACS{\par\leavevmode\hbox {\it PACS:\ }}
\newcommand{\fat}[1]{\mbox{\boldmath $ #1 $\unboldmath}}

\title{The BCS model  and the off shell Bethe ansatz 
for vertex models}
\author{Luigi Amico$^{[1]}$, G.  Falci$^{[1,2]}$, and Rosario  
Fazio$^{[1]}$}
\address{[1] Dipartimento di Metodologie Fisiche e Chimiche (DMFCI), 
	Universit\'a di Catania, viale A. Doria 6, I-95125 Catania, Italy\\
	Istituto Nazionale per la Fisica della Materia, 
	Unit\'a  di Catania, Italy}
\address{[2] Laboratoire d'Etudes des Propri\'et\'es Electroniques
	des Solides, Centre National de la Recherche Scientifique, 
	BP 166, 38042 Grenoble Cedex
	9, Grenoble, France} 
\date{\today}
%{\small 
%\begin{center}
%{\bf \abstractname}
%\end{center}
\begin{abstract}
We study the connection between the BCS pairing model and 
the inhomogeneous vertex model. The two spectral problems  
coincide  in the quasi--classical limit of the off--shell Bethe 
Ansatz of the disordered six vertex model. 
The latter problem is transformed into an auxiliary 
spectral problem which corresponds to  the diagonalization of
the integrals of motion of  the BCS model. 
A generating  functional whose quasi classical expansion leads 
to the constants of  motion of the BCS model 
and in particular the Hamiltonian, is identified.
\end{abstract}

\PACS 03.65.Fd,  74.20.Fg

{\it keywords:} Integrable models; BCS Superconductivity.
%\maketitle
%} 
%%%%%%%%%%%%%%%%%%%%%%%%%%%%%%%%%%%%%%%%%%%%%%%%%%%%%%%%%%%%%%%%%%%
%
\section{Introduction}
One of the most successful models of interacting electrons
is the BCS model of pairing~\cite{BCS}. Originally proposed  to describe 
properties of superconductors~\cite{TINKHAM}, the pairing idea 
has been applied to a large variety 
of physical systems in nuclear physics~\cite{IACHELLO94} and in 
QCD~\cite{RISCHKE}.  Recent experiments in metallic 
nanoparticles~\cite{EXP} have renewed the interest in the problem of 
pairing  correlations in mesoscopic systems~\cite{MASTELLONE}. 
The BCS Hamiltonian consists of a kinetic and an interaction term which 
describes the attraction between electrons in time reversed states
\begin{equation}
H  = \sum_{j=1 \atop \sigma= \up, \dwn}^\Omega 
\varepsilon_{j\sigma} c_{j \sigma}^\dagger c_{j \sigma}
  - g  \sum_{j, j'=1}^\Omega   c_{j\up}^\dagger c_{j \dwn}^\dagger 
c_{j' \dwn} c_{j' \up}  \qquad.
\label{pairing}
\end{equation}
The quantum number $j\in \{ 1, \dots, \Omega\}$, $\sigma \in \{\up,\dwn \}$  
labels a shell of doubly degenerate time reversed states of energy 
$\epsilon_j$;
$c_{j,\sigma}$ and $c_{j,\sigma}^\dagger $ are the corresponding 
electronic operators; $g$ is the  BCS coupling
constant. The low energy properties associated to this model are universal
functions of the ratio between two energies, the single particle average level 
spacing and the BCS gap~\cite{MASTELLONE}. 
\\
Various exact results have been obtained for the BCS Hamiltonian.
\\
In the limit $g \to \infty$ the exact eigenvalues and eigenstates 
can be  found (see for example Ref.~\cite{KLEINERT})   
and the integrals of motion 
are Gaudin  Hamiltonians~\cite{GAUDIN}. 
An important consequence of  the relation  with the 
isotropic Gaudin magnet (discussed in  Appendix A), 
is that the  Quantum Inverse Scattering Method 
(QISM)~\cite{KOREPIN-BOOK} for the $g \to \infty$ BCS model can be related with the 
QISM for the Gaudin model~\cite{SKLYANIN}. 
The  $g\to \infty $ BCS model can be also  related to the 
inhomogeneous 
vertex models~\cite{BAXTER,NOTE-VERTEX}. 
\\
Much less work has been done for finite $g$.
The exact solution was found by Richardson and Sherman 
(RS)~\cite{RICHARDSON}   
and independently by Gaudin~\cite{GAUDIN-BCS} by means of  the 
Bethe Ansatz (BA) technique. 
Approximate expression of correlation functions were found 
in the Ref.~\cite{RICHARDSON-CORRELATION}.
More recently, the integrals of motion of the BCS model  were 
obtained~\cite{CAMBIAGGIO,SKLYANIN} and  were diagonalized 
by means of the algebraic BA. 
\\
In this work we  show that for finite $g$
the BCS model    is connected to a 
disordered six vertex model
through the  {\it Off--Shell} BA (OSBA) introduced by Babujian {\it et al.}
in Refs.~\cite{BABUJIAN,NOTE-OSBA}. In this framework, 
the known connection between 
the isotropic Gaudin models and the inhomogeneous vertex model 
is obtained as the  {\it Mass--Shell} limit which corresponds to 
$g \to \infty$. 
\\
A strong hint towards our result is provided by a recent work by 
Sierra~\cite{SIERRA} who has shown  connection between 
the BCS pairing model and a $su(2)_c$ 
Wess-Zumino-Novikov-Witten Conformal Field Theory (CFT), in the singular 
limit when the central charge is infinite;   
the RS wave functions solve the 
Knizhnik--Zamolodchikov equations for the CFT correlation functions.
The results of Sierra are indeed related to the connection existing between   
models in statistical mechanics and correlations functions
of a suitable CFT established through the OSBA.
In fact the solution of the  quasi classical  
OSBA equations is equivalent to solve the Knizhnik--Zamolodchikov 
equations~\cite{RESHETKIN,K-Z} and in particular 
the  quasi-classical OSBA equations for the vertex models generate 
the correlators of the $su(2)$ Wess-Zumino-Novikov-Witten CFT.

The paper is organized  as follows.
In  section II we review the exact solution of the pairing 
model.
In section III the quasi-classical expansion of the OSBA 
of the disordered six vertex model is identified as the diagonalization 
of  the BCS model. Section IV is devoted to the conclusions.  
The  connection between the diagonalization of the pairing model for 
infinite  pairing coupling constant
$g$ and  the diagonalization of Gaudin magnet is reviewed in Appendix A.
In the Appendix B we summarize the QISM of the inhomogeneous vertex model.

%%%%%%%%%%%%%%%%%%%%%%%%%%%%%%%%%%%%%%%%%%%%%%%%%%%%%%%%%%%%
\section{The exact solution of the pairing Hamiltonian}

In this section we review the exact 
solution~\cite{RICHARDSON,GAUDIN-BCS} 
of the BCS 
model (Eq.~(\ref{pairing})) and
the formulation of its integrability.
Due to the form of the pairing interaction 
in Eq~(\ref{pairing}), single occupied states are 
frozen and 
we can focus 
on scattering of pairs.
The Schr\"odinger equation for a state of $N$ Cooper pairs 
\begin{equation}
 H|N \rangle 
={\cal E} |N \rangle \;; 
\label{diagonal-pairing} 
\end{equation}
has the solution~\cite{RICHARDSON,GAUDIN-BCS} 
\begin{eqnarray} 
\label{BA-state} 
|N \rangle &=& \prod_{\alpha = 1}^N \sigma^+(e_\alpha, \varepsilon) |0 \rangle 
\qquad \quad ; \quad
\sigma^+(e_\alpha, \varepsilon)
\;:=\;
\sum_{j=1}^\Omega \frac{\sigma^\dagger_j}{2 \varepsilon_j - e_\alpha} 
\nonumber \\
{\cal E} &=& \sum_{\alpha =   1}^{N} e_\alpha
\end{eqnarray}
The operators  
$
\sigma^-_j := c_{j,\dwn} c_{j,\up}$, $\sigma^+_j =(\sigma^-_j)^\dagger$ 
and  
$\sigma^z_j := (c^\dagger_{j,\up} c_{j,\up}+ c^\dagger_{j,\dwn} 
c_{j,\dwn}-1)/2$ 
realize $su(2)$  in the lowest representation.
The vacuum state is the highest weight vector of $su(2)$: 
$|0\rangle:=|1/2,-1/2\rangle$.  
The operators $\left\{\sigma^\pm (e_\alpha, \varepsilon), 
\sigma^z(e_\alpha, \varepsilon) \right\}$ 
generate (for generic $e_\alpha$)  the Gaudin algebra ${\cal G}[sl(2)]$ 
(see the Appendix A and Eqs.~(\ref{s-gaudin-algebra}) of the next section).
The energy ${\cal E}$ is 
given in terms of the spectral parameters $e_\alpha$ 
which satisfy the algebraic equations~\cite{RICHARDSON}
\begin{equation}
\frac{1}{g } + \sum_{\beta=1 \atop \beta\neq \alpha}^{N} \frac{2}{ e_\beta- e_\alpha} 
- \sum_{j=1}^\Omega \frac{1}{2 \varepsilon_j - e_\alpha} =0\; , 
\quad \alpha = 1, \dots , N  \; .\label{re}
\end{equation}
The method employed by RS has analogies  with the  coordinate  BA  
technique. In fact, in the coordinate BA the ansatz functions are plane waves 
(describing free particles) modified to include the interaction.
In the RS solution  the ansatz functions   
are the solutions of the model when pairs of time-reversed electrons are 
treated as bosons;  these functions are modified because
Cooper pairs behave as hard-core bosons.
In both the RS and the BA procedures the modification enters  the  set of 
the algebraic equations for the rapidities (Bethe equations) 
parameterizing the eigenvalues of the Hamiltonian.

By using the spin realization of pair operators $\{\sigma^z_j,\sigma^\pm_j\}$, 
the pairing Hamiltonian can be written   
as a quantum spin model with 
long range interaction  in a non uniform
fictitious magnetic field, given by  $\varepsilon_j$
\begin{equation}
 H=  \sum_{j=1}^\Omega  2 \varepsilon_j \sigma^z_j - {{g}\over{2}}
\sum_{j,l=1}^\Omega( \sigma^+_l \; \sigma^-_j + \sigma^+_j \; \sigma^-_l ) +const\; . 
\label{xy}
\end{equation} 
Cambiaggio {\it et. al.}~\cite{CAMBIAGGIO} found that the integrals of motion 
$\tau_j$ 
of this 
model, if
$\varepsilon_j \neq \varepsilon_l$,  $\forall j \neq l$
have the form
\begin{equation}
\label{crs}
\tau_j \;=\; {{1}\over{g}} \sigma^z_j -  \; \Xi_j \;,
\label{integrals} 
\end{equation}
and satisfy the commutation relations 
$[H, \tau_j] =  [\tau_j, \tau_l] = 0$, $\forall j, l \in \{1, \dots, \Omega\}$.  
The operators $\Xi_j$ in Eqs.~(\ref{integrals}) 
are spin--$1/2$ Gaudin Hamiltonians~\cite{GAUDIN}
\begin{equation}
\Xi_j \;:=\; \sum_{l=1 \atop l\neq j}^\Omega 
\frac{ {\fat{\sigma }}_j \cdot {\fat{\sigma}}_l }
	{ \varepsilon_j- \varepsilon_l} \;. 
\label{gaudin-op}
\end{equation}
The commuting operators $\tau_j$  were also found by Sklyanin~\cite{SKLYANIN} 
by taking the quasi-classical limit of the monodromy matrix of the 
inhomogeneous  vertex model {\it twisted} by a term proportional to 
$\sigma^z_j/g$.
The pairing Hamiltonian can be 
expressed as function of the  integrals of motion as
\begin{equation}
{{1}\over{g^3}} \, H \;=\; {{1}\over{g^2}} \, 
\sum_{j=1}^\Omega  \,2 \varepsilon_j \,\tau_j \,+\,  
\sum_{j,l=1}^\Omega \tau_j  \, \tau_l  \,+ \,const.
\label{integrals-pairing}
\end{equation}
In the limit $g \to \infty$ the problem is 
equivalent to the diagonalization of (all) the Gaudin Hamiltonians
(see  the Appendix A for details).

\section{The OSBA of the inhomogeneous vertex model and the pairing model}

Vertex models are two dimensional   classical 
models which were solved  
long ago by inverse methods~\cite{BAXTER}. 
Generalizations to $su(2)$ higher  
representations
and to include disorder were intensively 
studied~\cite{INHOMOGENEOUS,DEVEGA,KULISH}.

\begin{figure}
\centerline{\psfig{figure=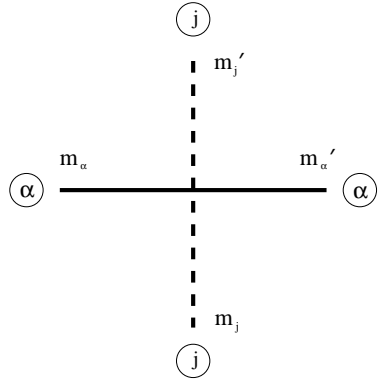,height=5cm}}
\caption{The vertex $(\alpha,j)\in \Lambda$. Disorder $z_j$ is distributed 
along the vertical lines. It is due to both the spin inhomogeneity and impurities $I_j$ distributed in the lattice:
$z_j \equiv z(S_j)+z(I_j)$}
\label{vertex}
\end{figure}

In this section we introduce a vertex model in which the inhomogeneity is
due to the combination of  given (see below) disordered  distribution 
of both the spin  and the impurities in the lattice. Then  we 
apply the scheme developed by 
Babujian {\it et. al.}~\cite{BABUJIAN}
to relate this inhomogeneous  vertex models to the BCS model 
 Eq.~(\ref{pairing}).
\\
The  model is defined in the following way. 
On the edges of the square lattice $\Lambda: N_v\times N_h$   
($N_v$ columns  and $N_h$ rows) are arranged 
$N_v+1$ types of spin variables.
On horizontal edges  (labelled by $\alpha=1\dots N_h$) are arranged the spins $\sigma$ 
taking spin projection $m_\alpha\in \{\pm 1/2\}$.
On the columns  (labelled by $j=1\dots N_v$)  the spin variables $S_j$ can 
take any value $m_j \in \{-s_j,\dots ,+s_j\}$
of the $s_j$-th representation of $su(2)$. 
The partition function is restricted to 
configurations for which 
an even number of spins are into (or out of) each lattice site (vertex); 
configurations in which the four spins are all in or all out are excluded 
(ice rule). 
The ``scattering''  between spin states 
$(m_\alpha,m_j)\,  \to\,  (m_\alpha',m_j')$ 
of vertex  $(\alpha, j)$ (see Fig.~\ref{vertex}) have weights fixed by the 
universal matrix elements 
$R_{m_\alpha,m_\alpha'}^{m_j,m_j'}(\lambda -z_j)$ where 
$\lambda$ is the spectral parameter.
The quantities $ z_j$ (see also Ref.~\cite{DEVEGA})  shift spectral parameters 
as  inhomogeneities  which we assume distributed only along the columns of $\Lambda$ (Fig.~\ref{vertex}).

The variables $z_j$ take into account of disorder induced  
by  the mixture of spin 
representations 
and/or  by  the actual distribution of impurities $I_j$ in $\Lambda$: 
$z_j \equiv z(S_j)+z(I_j)$. 
We assume that both $z(S_j)$ and $z(I_j)$ enter the universal matrix 
(see Eq.~(\ref{R-matrix})) in the same functional form.
The  
disordered six vertex model corresponds to the choice:    
$z(S_j)= 0$, $z(I_j)\neq 0$.
We impose periodic boundary conditions. 
\\
The transfer matrix 
$T(\lambda|{\bf z})$, where  
${\bf z}:=(z_1\dots z_{N_h})$,  can be expressed in terms of 
rational $R$--matrices  $R_X$, $X=\{\sigma, S\}$ (see Eq.~(\ref{R-matrix}))
which fulfill Yang Baxter relations 
(see the Appendix \ref{appendix-vertex}). This implies the  
integrability of the model:
$[T (\lambda|{\bf z}),T (\mu|{\bf z})]=0$. 

The application of the transfer matrix to the 
Bethe vector $\Phi (  \lambda_1 \dots \lambda_{N}|{\bf z})$ reads: 
\begin{eqnarray}
T(\lambda|{\bf z}) \Phi (  \lambda_1 && \dots \lambda_{N}|{\bf z})=
\Lambda ( \lambda, \lambda_1 \dots \lambda_{N}|{\bf z})
\Phi (  \lambda_1 \dots \lambda_{N}|{\bf z}) \nonumber \\
&&
-\sum_{\alpha=1}^{N} 
{{F_\alpha  }\over{\lambda - \lambda_\alpha}} 
\Phi_\alpha(\lambda_1..\lambda_{\alpha-1}, \lambda, \lambda_{\alpha+1}..\lambda_{N}|{\bf z}) \;, 
\label{OSBA}
\end{eqnarray} 
(for the explicit form of the quantities 
	$T(\lambda|{\bf z})$, $\Phi (  \lambda_1 \dots \lambda_{N}|{\bf z})$,
	$\Lambda ( \lambda, \lambda_1 \dots \lambda_{N}|{\bf z})$,  
	$F_\alpha $, and $\Phi_\alpha(\lambda_1..\lambda_{\alpha-1}, \lambda, \lambda_{\alpha+1}..\lambda_{N}|{\bf z})$ in (\ref{OSBA}),
	see Ref.~\cite{BABUJIAN}).
The condition for the diagonalization of $T$ (and of the constants of motion 
generated by  $T$) is that the spectral parameters
are chosen to  cancel the ``unwanted terms''(the second contribution to 
Eq.~(\ref{OSBA}))
in the spectral 
problem~(\ref{OSBA}); a sufficient condition is: $F_\alpha=0$ 
(algebraic BA equations).   
Such a condition  has   been termed as 
``Mass--Shell'' constraint~\cite{BABUJIAN} imposed on Eq.~(\ref{OSBA}).
The OSBA spectral problem, instead,  arises when the ``unwanted 
terms'' 
are considered in Eq.~(\ref{OSBA}); the spectral parameters obey a new set 
of equations called OSBA equations (see Eq.~(\ref{osbaeq}) below).
unfixed). The statement 
of the OSBA problem has been shown remarkable in the quasi--classical 
limit. In fact, it was discovered that the solutions of the 
quasi--classical 
OSBA problem satisfy the Knizhnik--Zamolodchikov 
equations~\cite{RESHETKIN,K-Z} for the $su(2)$ CFT. 
In the following we shall see how 
the quasi-classical limit of the OSBA problem for 
the disordered vertex model is solved by spectral parameters fulfilling 
the Eqs.~(\ref{re}). 

The quasi--classical limit of the vertex model is obtained  through 
the expansion of $R_X(x;\eta)$ in powers of  $\eta$
($R_X(x;0)$ is the identity). 
\\
Using the expressions for the monodromy, transfer, and universal 
matrices Eqs.~(\ref{R-matrix}), (\ref{monodromy}), the quasi--classical 
limit of the OSBA  Eq.~(\ref{OSBA}) reads
\begin{eqnarray}
\sum_{j=1}^{N_v} {{H_j}\over{\lambda-z_j}}\phi=h \phi + 
\sum_{\alpha=1}^{N} {{f_\alpha}\over{\lambda-\lambda_\alpha}}\phi_\alpha \;,
\label{quasi-int}
\end{eqnarray} 
up to  order  ${\cal O}(\eta^{N+2})$, where 
the explicit form of $h$ and $\phi_\alpha$ in  (\ref{quasi-int}) is given 
in the  Ref.~\cite{BABUJIAN}. 
By integrating Eq.~(\ref{quasi-int}) 
on a closed loop in the complex $\lambda$--plane 
encircling the pole $\lambda=z_j$
we obtain
\begin{equation}
H_j\phi=h_j \phi+ 
\sum_{\alpha=1}^{N} {{f_\alpha}\over{z_j-\lambda_\alpha}} S^+_j \phi_\alpha ' \; ,
\label{quasi}
\end{equation}  
\begin{eqnarray}
&&h_j=\sum_{l=1 \atop l\neq j}^{N_v}{{s_l s_j}\over{z_l-z_j}}-
\sum_{\alpha=1}^{N} {{s_j }\over{\lambda_\alpha-z_j}}\;,  \quad j=1\dots N_v \;, \label{energy}\\
&&f_\alpha=\sum_{\beta=1 \atop \alpha\neq \beta}^{N} {{1}\over{\lambda_\alpha-\lambda_\beta}}-
\sum_{j=1}^{N_v} {{s_j }\over{\lambda_\alpha-z_j}}\;, \quad \alpha=1\dots N \;.
\label{osbaeq}
\end{eqnarray}
The Bethe vectors in the quasi-classical limit are
\begin{equation}
\phi:= \prod_{\beta=1}^{N} S^+(\lambda_\beta,z)|vac\rangle\; ,\quad  
\phi_\alpha ':= \prod_{\beta=1 \atop \beta\neq \alpha}^{N} 
S^+(\lambda_\beta,z) |vac \rangle\; , 
\end{equation}
Here $|vac\rangle = 
\otimes_{j=1}^{N_v} 
|s_j,-s_j\rangle $, where 
$S_j^- |s_j,-s_j\rangle=0$, i.e.
$|vac\rangle$ is the highest weight vector in 
$\otimes_jsu(2)_{j}$. 
The three operators
$S^{\pm,z}(\lambda_\beta,z):=
\sum_{j=1}^{N_v} S^{\pm,z}_j /(z_j - \lambda_\beta)$ 
generate 
higher dimensional representations of the Gaudin algebra 
${\cal G}[sl(2)]$, given by (see also Eq.~(\ref{gaudin-algebra}))
\begin{eqnarray}
[ S^z(\lambda_\alpha, z),S^\pm(\lambda_\beta,z)] &=&
\pm {{S^\pm(\lambda_\alpha, z)-S^\pm(\lambda_\beta,z)}\over{\lambda_\beta-\lambda_\alpha}} 
\nonumber\\ 
\quad 
[ S^+(\lambda_\alpha, z),S^-(\lambda_\beta,z)] &=&
{{S^z(\lambda_\alpha, z)-S^z(\lambda_\beta,z)}\over{\lambda_\beta-\lambda_\alpha
}} \qquad .
\nonumber
\label{s-gaudin-algebra}
\end{eqnarray}

The ``Mass-Shell'' constraint: $f_\alpha=0$ corresponds  to 
the diagonalization of the Gaudin model (see Appendix A).

The solution of the spectral problem for the pairing model is 
recovered substituting
\begin{equation}
f_\alpha={{1}\over{2}} \left ( 
\sum_{j=1}^{N_v} {{1-2 s_j }\over{\lambda_\alpha-z_j}}+ {{1}\over{g}} \right )
\;,\quad \alpha=1\dots N \;, 
\label{osbaeq-pairing}
\end{equation}
in the left hand side of Eqs.~(\ref{osbaeq}). In fact, the resulting equations 
coincide with Eqs.~(\ref{re}). 
Substituting Eqs.~(\ref{osbaeq-pairing})  in (\ref{quasi}), 
and summing over index $j=1\dots N_v$ we obtain 
\begin{equation}
\sum_{\alpha=1}^{N} \left (\sum_{j=1}^{N_v} {{4 s_j-1}\over{\lambda_\alpha -
z_j}} +
{{1}\over{g}} \right ) \phi=0 \;, 
\label{spectral-osbae}
\end{equation}
where we have used the fact that:
$
\sum_{j=1}^{N_v} h_j=-\sum_{\alpha=1}^{N} \sum_{j=1}^{N_v} 
{s_j }/(\lambda_\alpha-z_j)
$.
Eq.~(\ref{spectral-osbae}) shows that, 
the OSBA spectral problem is transformed in a spectral problem 
involving only  diagonal matrix elements of suitably shifted 
(by $f_\alpha$) transfer matrix of the vertex model (in the quasi 
classical limit).
\\
Since the limit $g\rightarrow \infty$ should correspond to 
the same result of $f_\alpha\rightarrow 0$ for generic  $s_j$ (compare with 
Eqs~(\ref{ges})), we impose that the  distribution of spins $S_j$  through 
the lattice fulfills the condition 
\begin{equation}
\label{s-distribution}
\sum_{j=1}^{N_v} {{(1-2 s_j )}\over{(\lambda_\alpha-z_j)}}\equiv 0 \;.
\end{equation} 
In this case Eq.~(\ref{osbaeq-pairing}) reduces to
\begin{equation}
f_\alpha={{1}\over{2g}} 
\;,\quad \alpha=1\dots N \; .
\label{gaudin-conject}
\end{equation} 
We choose  $s_j=1/2 \quad \forall j$ in order to fulfill 
Eq.~(\ref{s-distribution}):  
the inhomogeneous vertex model becomes 
the disordered six vertex model since $z(S_j)= 0$ and $z(I_j)\neq 0$. 
This implies that 
$$
 H_j\equiv \Xi_j \; ,\, \makebox{and}
\quad \phi\equiv |N\rangle \; , $$
where: $N_v = \Omega$  
(compare with Eqs~(\ref{BA-state}), (\ref{gaudin-op})).
Eq.~(\ref{spectral-osbae}) can be transformed in the following eigenvalue  
equation:
\begin{eqnarray}
\label{eigenvalue-problem}
&&\sum_{j=1}^\Omega  \sum_{\alpha=1}^N \left (
{{-\sigma^z_j}\over{2\varepsilon_j -e_\alpha}}
 - {{1}\over{N}}\Xi_j \right ) \phi = \sum_{j=1}^\Omega  \sum_{\alpha=1}^N 
\tau_{j,\alpha} \phi  \;,  \\ 
&&
\tau_{j,\alpha}:=\left ({{1}\over{ \Omega g}}- 
{{1}\over{N}}h_j \right )
\; ,
\label{pairing-final}
\end{eqnarray}
where  Eqs.~(\ref{re}) (or 
(\ref{osbaeq}), (\ref{gaudin-conject}) ), (\ref{energy}),  and:
$
\sum_{\beta=1}^N \sigma^z(e_\alpha,\varepsilon)\phi=1/2
\sum_{\beta=1}^N \sum_{j=1}^\Omega 1/(e_\alpha-2\varepsilon_j)
\phi
$ have been used 
(the parameters in Eqs.~(\ref{energy})--(\ref{pairing-final}) 
are redefined as $ 
z_j\leftrightarrow 2 \varepsilon_j \; , 
\, \lambda_\alpha  \leftrightarrow e_\alpha\; $).
We point out that quantities in~(\ref{pairing-final})
are the eigenvalues of  operators  
$\tau_j$ in Eq.~(\ref{integrals})   for generic  $\Omega/ N$. 
At ``half filling'' $\Omega=2 N$ Eq.~(\ref{pairing-final})  reduces to
\begin{eqnarray}
\tau_{j,\alpha}
= {{1}\over{\Omega}} \left ({{1}\over{g}} -2 h_j \right )  
\;, 
\,  (j=1, \dots, \Omega) \; .\label{pairing-cambiaggio}
\end{eqnarray}
Equations~(\ref{pairing-final}), (\ref{pairing-cambiaggio}) coincide with 
those ones found by Sklyanin~\cite{SKLYANIN} and 
by Sierra~\cite{SIERRA}.
\\
The main result obtained in this paper is the 
connection between Eqs.~(\ref{quasi}) and 
Eqs.~(\ref{eigenvalue-problem}),(\ref{pairing-final}) through 
the Eq.~(\ref{osbaeq-pairing}).
The OSBA problem for the disordered six vertex model
(which does not account for  diagonalizing   the transfer matrix of the 
vertex model) reveals  the existence of a class of spectral problems 
(parameterized by $f_\alpha$)
which turns out to be diagonal on the quasi-classical Bethe vectors basis.
For $f_\alpha$ fixed by Eqs.~(\ref{gaudin-conject}) the 
diagonalization of  the BCS model  is obtained. 
\\ 
Furthermore, what we have discussed so far implies that 
the pairing Hamiltonian can be  
obtained from functionals of $\tau_j$ whose quasi-classical expansions 
have the following form:
\begin{eqnarray}
\label{transfer-pairing}
{\cal T} (e|{\bf z})= \sum_{a=0}^\infty 
\eta^{2a}  
e^{a-1}\left [{{1}\over{2 g^2}}+ \; \tau(e)\right ]^{a} \;, 
\end{eqnarray}
with 
\begin{equation}
\tau(e):=\sum_{j=1}^\Omega {{\tau_j}
\over{e-2 \varepsilon_j}} \; .
\end{equation} 
We point out that $[{\cal T}(e|{\bf z}),{\cal T}(e'|{\bf z})]=0\;, \; 
\forall e,e'$ 
since quantities $\tau_j$'s  commute each other. 
The residue in the poles $e =2 \varepsilon_j$ of the $\eta^2$ 
coefficient  provides  the 
integrals of motion $\tau_j$. The residue of the $\eta^4$ coefficient  reads  
(see Eq.~(\ref{integrals-pairing}))
\begin{equation}
 \sum_{j,l=1}^\Omega \tau_j \tau_l +
{{1}\over{g^2}}\sum_{j=1}^\Omega 2 \varepsilon_j \tau_j ={{1}\over{g^3}} H \quad . 
\label{transfer-hamiltonian}
\end{equation}

\section{Conclusions}
We have established a novel connection 
between the disordered six vertex model and the BCS model for generic $g$
through the OSBA procedure.  The BCS  model is diagonalized 
by the quasi classical limit of the OSBA equations of the disordered 
six vertex model. Retaining certain  
off diagonal terms of  the transfer 
matrix of the vertex model corresponds to  
the diagonalization of the  integrals of motion of 
the pairing model for finite $g$. 
The ``mass shell'' condition (and then the diagonalization 
of the quasi classical transfer matrix  of the vertex model) reproduces the 
limit  $g\rightarrow \infty$; the corresponding problem
is the Gaudin spectral problem. 
\\
The integrals of motion of the BCS model coincide with 
the integrals found by Sklyanin\cite{SKLYANIN} by considering 
a twist in  the monodromy  matrix of the 
vertex model (see Eq.~\ref{monodromy}). The algebraic equations  
which diagonalize these  integrals of motion via  algebraic  BA 
(namely via the  Mass Shell BA procedure) coincide with the 
Richardson's equations. 
The present study shows that the Sklyanin's procedure  
produces the same results
of  the Off Shell BA procedure applied to the {\it untwisted} monodromy 
matrix.

The existence of the relation between BCS model and quasi-classical 
vertex models, found in the present paper,
is consistent with the correspondence 
between CFT and  the BCS  model recently found by Sierra~\cite{SIERRA}.

Eqs.~(\ref{re}) were already 
conjectured by Gaudin (see formulas (5.15), (5.16) of Ref.~\cite{GAUDIN})
as connected (through Jacobian of certain matrices) 
with the norms of the Bethe vectors, 
$\rm{det}(\partial f_\alpha/\partial e_\beta) \sim \| \phi \|$  
(see Refs.~\cite{RICHARDSON-CORRELATION},~\cite{KOREPIN-GAUDIN}).
In this work we  have shown  that the Jacobian is connected
with OSBA of the vertex model. 
This might be useful to compute norms 
(and scalar products) and, then, to express 
the correlation functions of the BCS model as suitable determinants.  
This  exact calculation is our  major task in the future.

\ack
We thank G. Sierra for constant and invaluable help since the early stages of 
this work. A. Osterloh is acknowledged for very useful discussions and for a 
critical reading of the manuscript. We thank F. Dolcini and  G. Giaquinta 
for discussions.
We acknowledge the financial support of INFM-PRA-SSQI and the
European Community (Contract FMRX-CT-97-0143).

\begin{appendix}

\section{The pairing model  and the Gaudin spectral problem}

In this Appendix we discuss the connection between the pairing model 
and the Gaudin model. 
\\
The limit $g\rightarrow \infty$ of the  constants of motion 
$\tau_j$~(\ref{crs}) 
coincides with Hamiltonians $\Xi_j$. Since Eq.~(\ref{integrals}), 
the spectrum of the pairing problem coincide with that one of 
the Gaudin  magnet~\cite{GAUDIN}:  
$\, \Xi(u):=\sum_{j = 1}^{\Omega} \Xi_j/(u-2 \varepsilon_j)$
($u$ is a complex parameter). 
The total energy is: $h(u):= \sum_{j = 1}^{\Omega}h_j/(u-2 \varepsilon_j) $ 
($h_j$ is fixed by Eq.~(\ref{energy})).
The Bethe vectors
of the Gaudin and the pairing problems coincide formally for any $g$ since
operators 
$(\sigma^\pm(u, \varepsilon), \sigma^z(u, \varepsilon))$ 
in~(\ref{BA-state}) 
generate 
the Gaudin algebra ${\cal G}[sl(2)]$ in the lowest representation:
\begin{eqnarray}
[ \sigma^z(u, \varepsilon),\sigma^\pm(w,\varepsilon)]&&=
\pm {{\sigma^\pm(u, \varepsilon)-\sigma^\pm(w,\varepsilon)}\over{w-u}} \; ,\\ \nonumber
\quad 
[ \sigma^+(u, \varepsilon),\sigma^-(w,\varepsilon)] &&=
{{\sigma^z(u, \varepsilon)-\sigma^z(w,\varepsilon)}\over{w-u
}} \; ,
\label{gaudin-algebra}
\end{eqnarray}
where $\sigma^z(u, \varepsilon):=\sum_{j=1}^\Omega \sigma^z_j /(2 \varepsilon_j - u)$. 
\\
However, the spectral parameters 
entering the eigenvectors of the two models satisfy, 
for generic $g$, different equations (compare with  Eqs.~(\ref{re}))
\begin{equation}
 \sum_{\beta=1 \atop \beta\neq \alpha}^{N} \frac{2}{ e_\beta- e_\alpha} -
 \sum_{j=1}^\Omega \frac{1}{2 \varepsilon_j - e_\alpha}=0 \; . 
\label{ge}
\end{equation}
We point out that~Eqs.~(\ref{ge}) are  the limit 
$g\rightarrow \infty$ of Eqs.~(\ref{re})
for the pairing model. 
In this limit 
the two models have the same eigenvectors 
(see Eqs~(\ref{BA-state}), (\ref{re}), (\ref{ge})). 
\\
Thus the diagonalization 
of the Gaudin model is equivalent to the diagonalization of the BCS model for $g \rightarrow \infty$.
\\
The limit of large $g$ in Eq.~(\ref{integrals-pairing})  gives: 
\begin{eqnarray}
H\approx &&-g\sum_{j,l=1}^\Omega 
{{1}\over{\varepsilon_j-\varepsilon_l}}
\left [ 
({\varepsilon_j+\varepsilon_l}) \sigma^z_j \; \sigma^z_l 
+ \varepsilon_j \sigma^+_j \; \sigma^-_l \right ] \nonumber \\
&&\equiv  
-{{g}\over{2}}\sum_{j,l=1}^\Omega \sigma^+_j \; \sigma^-_l \;, 
\end{eqnarray} 
which (consistently) reproduce the Hamiltonian~(\ref{pairing}) for large 
$g$.
\\
The QISM was applied to Gaudin magnet 
for generic spin ${\bf S}_j$ in Ref.~\cite{GAUDIN,GAUDIN1}. 
In this case the Gaudin Hamiltonians are 
$
H_l:= \sum_{j=1 \atop j\neq l}^\Omega 
{\bf S}_l \cdot {\bf S}_j /( \varepsilon_j- \varepsilon_l)
$. The spectral parameters obey
\begin{equation}
\label{ges}
 \sum_{\beta=1 \atop \beta\neq \alpha}^{N} \frac{1}{ e_\beta- e_\alpha} -
 \sum_{j=1}^\Omega \frac{s_j}{2 \varepsilon_j - e_\alpha}=0 \; , 
\end{equation}
where $-s_j$  corresponds to the  highest weight vector of ${\bf S}_j$.  

\section{Integrability of the inhomogeneous vertex model}
\label{appendix-vertex}

In this appendix we discuss the QISM of the inhomogeneous vertex 
model together with its quasi--classical expansion.

The universal matrix of the model reads
\begin{equation}
R_{X}(\lambda-z;\eta)=\id \otimes \id + 
f(\lambda-z,\eta)
\sigma \otimes X  \;, 
\label{R-matrix}
\end{equation}
where $f(x,\eta):= 2\eta/(\eta -2 x)$ depending 
on the arbitrary parameter $\eta \in \RR$.
The physical 
$R$--matrix  of the model corresponds to take 
$X\equiv S$ in (\ref{R-matrix}); the auxiliary one corresponds to 
$X\equiv \sigma$ and $z=0$. 
Both  these matrices fulfill 
homogeneous ($S\equiv \sigma$) Yang--Baxter relations; 
in addition they 
fulfill the the inhomogeneous ($S\neq \sigma$) 
one~\cite{INHOMOGENEOUS}:

\begin{eqnarray}
\label{2d-ybe}
R^{12}_\sigma (\lambda-\mu) 
R^{23}_S (\lambda-z)&&R^{12}_S (\mu - z)= \\
&&R^{23}_S (\mu-z) 
R^{12}_S (\lambda-z)R^{23}_\sigma (\lambda-\mu) \;,  \nonumber
\end{eqnarray}
where $R^{12}\doteq R\otimes \id$ and $R^{23}\doteq \id \otimes R$ act 
on the vector space $V_1\otimes V_2\otimes V_3$; $R^{12}$ and $R^{23}$ act  
as the identity 
in the vector spaces $3$ and $1$ respectively. Eq.~(\ref{2d-ybe}) is the sufficient 
condition for which the model, can be solved by diagonalizing 
the column to column (which is $2\times 2$ matrix in operators $S_j$) transfer 
matrix (instead of the row to row  one which is $(2s_j+1) \times (2s_j+1)$) 
obtained  through the trace in the two dimensional horizontal 
vector space (which is spanned by spins along the rows of 
$\Lambda$)  labelled by ``$(0)$'':
\begin{equation}
T (\lambda|{\bf z}):=tr_{(0)} J(\lambda| {\bf z}) \;. 
\label{monodromy}
\end{equation}
Twisted monodromy matrix is obtained by $J(\lambda| {\bf z})\to 
e^{a \sigma^z_{j}}J(\lambda| {\bf z})$.
The transfer matrices commute at different values of spectral parameters:
$[T (\lambda|{\bf z}),T (\mu|{\bf z})]=0$ (it proves the 
integrability of the  model) since the  monodromy 
matrix $J(\lambda| {\bf z}):=\prod_{j=N_v}^1 R^{(0)}_{S_j}(\lambda -z_j)$ 
satisfies: 
$
R_\sigma(\lambda-\mu) [J(\lambda| {\bf z}) \otimes J(\mu| {\bf z})]= 
[J(\mu| {\bf z}) \otimes J(\lambda| {\bf z})] R_\sigma(\lambda-\mu)
$ (induced by Eq.~(\ref{2d-ybe}).
The matrices $R^{(0)}_{S_j}$ and  $R_\sigma$ are 
\begin{equation}
R^{(0)}_{S_j}(x,\eta):= \left(
\begin{array}{cl}
\id+f(x,\eta) S^z_j & \quad f(x,\eta) S^-_j  \\
f(x,\eta) S^+_j  & \id  -f(x,\eta) S^z_j 
\end{array}
\right) \; ,
\end{equation}
\begin{equation}
R_\sigma(\lambda-\mu,\eta):= \left(
\begin{array}{cccccccc}
1 &\ &   0  &\ & 0  &\ & 0  \\
0 &\ &   c  &\ & b  &\ & 0   \\
0 &\ &   b  &\ & c  &\ & 0   \\
0 &\ &   0  &\ & 0  &\ & 1
\end{array}
\right) \; ,
\end{equation}
where: $b(\lambda-\mu):=\eta/(\eta-\lambda-\mu)$, 
$c(\lambda-\mu):=\lambda-\mu/(\lambda-\mu-\eta)$ 
(note that $R_\sigma(\lambda-\mu,\eta)$ is $z$--independent).

In the quasi--classical limit, the system generates a 
hierarchy of integrable  systems in 
the quasi--classical limit since
\begin{equation}
\sum_{a=b+c=0}^\infty [T_b (\lambda|{\bf z}),T_c (\mu|{\bf z})]=0 
\; , 
\label{hierarchy}
\end{equation}
where we have used the $\eta$--expansion of the transfer matrix:
$T (\lambda|{\bf z})=\sum_{a=0}^\infty \eta^a T_a (\lambda|{\bf z})$
(the sum in Eq.~(\ref{hierarchy}) is meant on ordered 
partitions of $a$ including $b\,  \vee\, c=0$).
Up to   order $\eta^2$, the transfer matrix reads
\begin{equation}
T (\lambda|{\bf z})=2 \id +2 \eta^2 \sum_{j=1}^{N_v} {{H_j}\over{\lambda-z_j}} \; ,
\label{transf}
\end{equation}
where
the Hamiltonians $H_j$ in Eq.~(\ref{transf}) 
are  
\begin{equation}
H_j= \sum_{l=1 \atop l\neq j}^{N_v} \frac{ {\bf S}_l \cdot {\bf S}_j }{ z_j- z_l} ,\;\;
(j=1, \dots, N_v) \; .
\label{iso-gaudin} 
\end{equation}
The transfer matrix~(\ref{transf}) coincide with the 
quasi-classical  expansion of the twisted Gaudin model's transfer matrix 
(see formula (1.16) of the Ref.~\cite{SKLYANIN}).

\end{appendix}

%%%%%%%%%%%%%%%%%%%%%%%%%%%%%%%%%%%%%%%%%%%%%%%%%%%%%%%%%%%%%%%%%%%%%%%%%%%%%%%
%%%%%%%%%%%%%%%%%%%%%%%%%%%%%%%%%%%%%%%%%%%%%%%%%%%%%%%%%%%%%%%%%%%%%%%%%%%%%%%


\begin{thebibliography}{100}
\bibitem{NOTE-VERTEX}The inhomogeneous vertex models 
	are related to  the 
	Gaudin models~\cite{SKLYANIN,BABUJIAN}:
	the BA equations of the Gaudin model can be obtained by 
	taking the quasi-classical limit of the BA equations of vertex models.
\end{thebibliography}
\end{document}